%% file: main.tex
\documentclass[10pt, conference]{IEEEtran}  

\IEEEoverridecommandlockouts                              

\usepackage{lipsum} 
\usepackage{fancyhdr}
\usepackage{graphicx}
\usepackage{cite}
\usepackage{mathtools, cuted}
\usepackage{booktabs}
\usepackage{threeparttable}
\usepackage{comment}
\usepackage{authblk}
\usepackage{graphics}
\usepackage{threeparttable}
\usepackage[table,xcdraw]{xcolor}
\usepackage{multicol}
\usepackage{multirow}
\newcommand{\name}{\texttt{SALTY}}
\usepackage{xcolor}
\usepackage[linesnumbered,ruled,vlined]{algorithm2e}
\SetKwRepeat{Do}{do}{while}%
\usepackage{amsmath,amssymb}
\usepackage{caption}
\usepackage{paralist}
\usepackage{dblfloatfix}
\usepackage{subcaption}
\usepackage{setspace}
\usepackage{wrapfig}
\usepackage[hyphens]{url}
\usepackage{amsthm}
\usepackage{relsize}
\usepackage{amssymb}
\usepackage{pifont}
\usepackage{float}
\usepackage{graphicx}

\usepackage[table,xcdraw]{xcolor}
\newcommand{\xmark}{\ding{55}}%

\begin{document}

\title{SALTY: Explainable Artificial Intelligence Guided Structural Analysis for Hardware Trojan Detection}

\author[1]{Tanzim Mahfuz}
\author[2]{Pravin Gaikwad}
\author[1]{Tasneem Suha}
\author[2]{Swarup Bhunia}
\author[1]{Prabuddha Chakraborty}
\affil[1]{Department of Electrical \& Computer Engineering, University of Maine, Orono, ME, USA}
\affil[2]{Department of Electrical \& Computer Engineering, University of Florida, Gainesville, FL, USA}

\date{}

\maketitle
\thispagestyle{fancy}

\fancyhf{}

\fancyfoot[C]{\large \textbf{ \textcolor{blue}{© This paper has been accepted for publication at 43\textsuperscript{rd} IEEE VLSI Test Symposium (VTS), 2025}}}
\renewcommand{\footrulewidth}{0.4pt}
\renewcommand{\headrulewidth}{0pt}
\renewcommand{\footrulewidth}{0pt}


\begin{abstract}
Hardware Trojans are malicious modifications in digital designs that can be inserted by untrusted supply chain entities. Hardware Trojans can give rise to diverse attack vectors such as information leakage (e.g. MOLES Trojan) and denial-of-service (rarely triggered bit flip). Such an attack in critical systems (e.g. healthcare and aviation) can endanger human lives and lead to catastrophic financial loss. Several techniques have been developed to detect such malicious modifications in digital designs, particularly for designs sourced from third-party intellectual property (IP) vendors. However, most techniques have scalability concerns (due to unsound assumptions during evaluation) and lead to large number of false positive detections (false alerts). Our framework (\name) mitigates these concerns through the use of a novel Graph Neural Network architecture (using Jumping-Knowledge mechanism) for generating initial predictions and an Explainable Artificial Intelligence (XAI) approach for fine tuning the outcomes (post-processing). Experiments show $>98\%$ True Positive Rate (TPR) and True Negative Rate (TNR), significantly outperforming state-of-the-art techniques across a large set of standard benchmarks.  


\end{abstract}

\begin{IEEEkeywords}
Hardware Trojans, Hardware Security and Trust, Explainable Artificial Intelligence
\end{IEEEkeywords}


\section{Introduction}
The global distributed supply chain has a massive economic advantage but suffers from security and trust concerns. Most modern semiconductor companies operate in a fabless model, including economic giants such as NVIDIA, Apple, AMD, Qualcomm, and Broadcom that are driving the current AI wave. Most small semiconductor business are following the same example set by these larger companies. Most of these companies often acquire/license application-specific design blocks (also called IPs) from third-party vendors around the world, integrate these IPs into their core design, and send the final design to other entities for fabrication/testing/packaging. This distributed model opens up the system to a variety of threat vectors such as hardware Trojans, intellectual property theft, and privacy compromise through reverse engineering. 

In this study, we focus on hardware Trojans which are intentional malicious modifications in digital designs that can lead to certain non-characteristic behavior of the system in-field \cite{derms}. Such attacks can be carried out by malicious third-party IP vendors (3PIP), rogue employees in the design house, and an untrusted foundry. Typically hardware Trojans have a very small footprint (compared to the original design) in terms of area/power/delay and are rarely triggered during the system testing phase. Hence, it is a difficult task to detect these modifications in massive digital designs. Recent works such as VIPR \cite{gaikwad2023hardware, hoque2018hardware} and TrojanSAINT \cite{lashen2023trojansaint} have attempted to detect these Trojans in digital designs using different Artificial Intelligence (AI) techniques with varying degrees of success. VIPR suffers from low True Positive Rate (TPR) while TrojanSAINT has massive instability issues (performs poorly on certain designs). The post-processing techniques introduced in VIPR are inspiring, but lack dynamism and flexibility. Other techniques such as GNN4Gate \cite{cheng2022gnn4gate}, FAST-GO \cite{imangholi2024fast}, NHTD \cite{hasegawa2023node} have major scalability/practicality concerns because of unsound evaluation choices (trained and tested on similar benchmark structures). 

In this work, we propose \textbf{SALTY} (\textbf{S}tructural \textbf{A}I for Exp\textbf{l}ainable \textbf{T}rojan Anal\textbf{y}sis), a novel hardware Trojan detection framework (golden free) that attempts to mitigate the concerns and shortcomings of existing frameworks. \name~ utilizes a novel Graph Neural Network (GNN) with Jumping Knowledge (JK) mechanism while leveraging design structural features similar to SAIL \cite{chakraborty2021sail} (a reverse engineering framework). The JK-mechanism reduces the risk of over-smoothing by aggregating hidden layer embeddings. This allows the model to attain a much higher TPR and TNR compared to other state-of-the-art GNNs. Artificial Intelligence (AI) and statistical models are often not fully aware of the domain constraints and make irrational decisions (hallucination) leading to poor performance. To tackle this concern, we utilize explainability techniques to draw insights about when and how our AI models fail to make correct decisions. This insight allows us to build a dynamic post-processing module that can further enhance the performance of \name. We evaluate our framework on a large set of standard TrustHub benchmarks ($>15$) and compare the results against seven other state-of-the-art (SOTA) Trojan detection techniques. We observe significant performance improvements (TPR/TNR combined) and enhanced visibility into the Trojan detection process (via explainability).   
\input{fig_latex/methodology}
In particular, we make the following contributions:
\begin{enumerate}
    \item Designed a novel Graph Neural Network (GNN) that utilizes Jumping Knowledge (JK) mechanism for efficient hidden layer embedding aggregation.
    \item Developed a novel explainability guided dynamic post-processing module that can boost Trojan detection performance by removing hallucinations.
    \item Implemented the \name~framework as a highly parameterized tool and integrated it into the EDA flow. 
    \item Robust generalization in hardware Trojan detection, enabling the identification of Trojans in entirely unseen benchmarks without prior exposure to similar structures.
    \item Extensively evaluated \name~with $>15$ benchmarks against seven SOTA Trojan detection frameworks.
\end{enumerate}


\section{Related Works}\label{sec:background} 


A hardware Trojan (HT) is a malicious alteration or addition to a hardware design during its creation or assembly. These Trojans compromise the hardware by creating vulnerabilities or unintended functions, ranging from small circuits leaking data to system-disrupting modifications. Their stealth and dormancy make detection and mitigation difficult. 
Detecting hardware Trojans (HTs) remains a crucial research focus, prompting the development of various frameworks and methods to tackle this issue. In this section, we examine significant progress made in the domain, particularly emphasizing machine learning and graph-based approaches, and discuss their advantages and drawbacks. 


Various frameworks that leverage machine learning have been developed to detect hardware Trojans, each offering its advantages and drawbacks. VIPR \cite{gaikwad2023hardware} improves detection accuracy without relying on golden models through pseudo-self-referencing and post-processing, yet it is limited by its dependence on synthetic Trojans and feature selection. A hybrid XGBoost framework \cite{li2020xgboost} integrates static and dynamic analysis but faces issues with scalability and generalizability due to computational demands and dataset reliance. Graph Neural Networks (GNNs) effectively represent circuits as graphs, facilitating accurate detection. TrojanSAINT \cite{lashen2023trojansaint} breaks circuits into subgraphs to extract features, although it experiences challenges with reduced accuracy and scalability. GNN4Gate \cite{cheng2022gnn4gate} facilitates automatic Trojan detection without the need for golden models; however, it necessitates the manual design of various features, which complicates its application to real-world scenarios. GNN4HT \cite{chen2024gnn4ht} expands detection to gate-level and RTL designs, but is hampered by high interclass similarity and limited generalization due to insufficient data set augmentation. FAST-GO \cite{imangholi2024fast} employs Graph Convolutional Networks (GCNs) enhanced with improved features and dynamic thresholds for scalable detection. R-HTDetector \cite{hasegawa2022r} boosts robustness through adversarial training. NHTD-GL \cite{hasegawa2023node} automates feature extraction to detect threats at the node level. These methods operate at the net level and encounter difficulties with imbalanced datasets and feature initialization, while also lacking the ability to localize Trojans. 

\input{Table/Data_Splitting}

\section{Motivations}\label{sec:motivation} 
\subsection{Generalization to Entirely Unseen Digital Designs}
A key motivation for our work is to achieve robust generalization in hardware Trojan detection across completely unseen designs. Therefore, a detection framework must be capable of identifying Trojans in completely new designs without having been trained or validated on similar structures. Table~\ref{table:data_splitting} highlights the summary of current approaches.
\subsection{Informed Post Processing - Minimize AI Hallucinations}
While techniques like VIPR \cite{gaikwad2023hardware} have investigated post-processing, such approaches are static in nature and fail to take the advantage of more intricate patterns. We hypothesize that explainability techniques \cite{xdfs,ribeiro2016should,arrieta2020explainable, doshi2017towards, guidotti2018survey} can provide deeper insight into AI models opening pathways for more robust post-processing that can foster trust and minimize the AI hallucinations.   




\section{Methodology}\label{sec:method}

Inspired by Sec.~\ref{sec:motivation}, we have developed the \name~framework which: (i) Integrates the Jumping Knowledge mechanism to boost performance on unseen designs; (ii) Utilize explainable AI for reducing AI hallucinations. The overview of the \name~is illustrated in Fig.\ref{fig:method}.
\input{Algo/datamodel}
\subsection{Knowledge Discovery \& Model Design}
Let's assume, a graph $G_r$ = $(V,E)$ is defined, where $V$ represents vertices and $E$ denotes hyperedges. Here, $V$: wires in a digital design, and $E$: connections between wires. Algo.~\ref{algo:Model Genesis} outlines the initial stages from graph generation to model creation using the Trojan inserted design ($\mathcal{D}$), ground truth ($\mathcal{T}_{lb}$) for wire labels, and locality ($\mathcal{L}$) as inputs. Locality ($\mathcal{L}$) takes care of the number of neighbors that needs to be collected from the sub-graph. In line 1, the \textit{graphify} constructs $G_r$, with connection details $E_d$ (edge-index), in line 2. Subsequently, all nodes ($\mathcal{D}_{wires}$) of the graph are gathered, and the structural features of each wire and the corresponding true label from ($\mathcal{T}_{lb}$) are extracted and organized into the dataset ($X\_data$, $Y\_data$)(lines 4-7). \name~employs local structural features of the sub-design, similar to the approach used in SAIL \cite{chakraborty2018sail} and SAIL-based works \cite{chakraborty2024learning}. The structural features of a wire describe its local placement and how it connects to other wires through the gates it interacts with. In a subgraph, we encode the connectivity using an adjacency matrix and represent gate types through one-hot encoding, similar to \cite{xdfs}. Fig.~\ref{fig:method} illustrates the structural features extraction process, showing how the gates are vectorized and how the dataset is created via feature extraction and labeling. We utilize the Breadth-First Search (BFS) algorithm to explore neighboring gates around a specific wire - the one for which we are extracting features and assigning a label. The dataset is used for  training and the trained model, $\mathbb{M}$ is returned in line 10. In this work, we have used Graph Attention Networks (GATs) \cite{velickovic2017graph} and Jumping Knowledge (JK) mechanism \cite{xu2018representation} which extend traditional GCN by introducing attention procedures, allowing the model to assign different importance to focus on the most relevant parts of the graph which is crucial for detecting subtle patterns associated with Trojans \cite{wu2020comprehensive}. Given a graph with node features $f_i$ for each node $i \in V$, a single GAT layer computes the updated node features $f_i'$ as follows:

\[
f_i' = \sigma \left( \sum_{j \in X(i)} \alpha_{ij} Wf_j \right)
\]
where: $X(i)$ is the set of neighboring nodes of node $i$, $W$ is a learnable weight matrix, $\sigma$ is a non-linear activation function, $\alpha_{ij}$ are the attention coefficients computed by:

\[
\alpha_{ij} = \frac{\exp \left( \text{LeakyReLU} \left( v^\top \left[ Wf_i \, \| \, Wf_j \right] \right) \right)}
{\sum_{k \in X(i)} \exp \left( \text{LeakyReLU} \left( v^\top \left[ Wf_i \, \| \, Wf_k \right] \right) \right)}
\]
Here, $v$ is a learnable weight vector, $\|$ denotes concatenation, and LeakyReLU is an activation function. The JK mechanism \cite{xu2018representation} combines information from different layers of GATs, addressing the issue of over-smoothing. Assuming we have node representations from $L$ layers, $\{f_i^{(1)}, f_i^{(2)}, \dots, f_i^{(L)}\}$, the JK mechanism aggregates these representations into a final node embedding $f_i^{JK}$ using:

\[
f_i^{JK} = \big\|_{l=1}^L f_i^{(l)}
\]
This approach retains information from all layers without loss. \name~consists of two sequential GAT layers aggregating with Jumping Knowledge which are fed into the $softmax$ activation layer. For the classification, we assign each node to the class corresponding to the highest predicted probability. 
\input{Algo/xai}
\subsection{Explainable AI: Post Processing Development}
 To introduce interpretability, we have used PyTorch Geometric's Explainer module, designed to identify influential nodes and edges in GNNs \cite{fey2019fast}. Within this module, we have leveraged the $Captum$-$Explainer$ algorithm \cite{kokhlikyan2020captum}. While other frameworks like SHAP \cite{lundberg2017unified} and LIME \cite{ribeiro2016should} are widely used for model interpretability, they may not effectively capture the complex relationships inherent in graph data. We utilize \textit{Integrated Gradients} \cite{sundararajan2017axiomatic}, a gradient-based attribution method from $Captum$-$Explainer$ that quantifies the contribution of each input feature to the model's prediction by integrating the gradients along a path from a baseline input to the actual input. For an input $\mathbf{x}$ and a baseline $\mathbf{x}'$, the attribution $\text{\textit{IG}}_{f_i}(\mathbf{x})$ for feature $f_i$ is calculated as:

\[
\text{\textit{IG}}_{f_i}(\mathbf{x}) = (x_{f_i} - x'_{f_i}) \times \int_{\alpha=0}^1 \frac{\partial F\left(x' + \alpha \times (x - x')\right)}{\partial x_{f_i}} \, d\alpha
\]
where $F$ is the model's output and $\alpha$ is a scaling factor between $0$ and $1$. Algo.~\ref{algo:XAI Development} explains the automated XAI process that enhances framework performance. Lines 1–3 extract the graph, edge-index, and structural features for each wire from the test design. Line 4 uses the trained model to predict non-Trojan and Trojan wires and lengths of the predictions stored in $x$ and $y$. The parameter $n$ specifies the number of top feature scores that the $Captum$-$Explainer$ module will provide. We eventually obtain top $n$ feature scores for both non-Trojan and Trojan nets from $F_0$ and $F_1$, respectively, which effectively distinguish between the two types of wires, in line 6. Aiming to detect Trojan nodes as accurately as possible, we take the top n average feature scores for both Trojan and non-Trojan wires (denoted as $\mathcal{K}$ and $\mathcal{V}$ array), in lines 7–9. The inputs $\mathcal{N}$ and $\mathcal{T}_{h}$ are crucial, as they determine when a non-Trojan node, initially predicted as non-Trojan (a false negative) should be reclassified as Trojan (a true positive). Specifically, $\mathcal{N}$ acts as a multiplier to check if summation of feature scores of a non-Trojan net is equal to or exceeds $\mathcal{N}$ times summation of the average feature scores of non-Trojan nets, line 11. Similarly, $\mathcal{T}_h$ represents the minimum allowable difference between summation of feature scores of a non-Trojan net and summation of the average feature scores of Trojan nets required for reclassification, line 12. If either condition is met, we adjust the prediction accordingly (prediction switched) in line 14. Finally, we calculate the True Positive Rate (TPR) and True Negative Rate (TNR) based on these updated predictions, as detailed in Sec.~\ref{sec:results}.
\input{Table/SAIL-HT_Locality}
\section{Results \& Analysis}\label{sec:results} 

\input{Table/SAIL-HT_main}
\input{fig_latex/explainability_Trojans}
\subsection{Experimental Setup}
To validate our proposed framework, we have established a comprehensive experimental setup to assess its effectiveness. In both algorithms, we set the parameter $\mathcal{L}$ = $7$. For algorithm~\ref{algo:XAI Development} specifically, we have used $\mathcal{N}$ = $2$, $\mathcal{T}_h$ = $0.1$, and $n$ = $10$ to inform decisions based on the XAI features. Our experiments utilize digital designs obtained from Trust-Hub \cite{trsuthub}. Regarding training parameters, we employed a weighted binary cross-entropy loss function with a batch size of $64$, the Adam optimizer, and learning rate $0.005$. $20\%$ of the training data was set aside for validation purposes. Thus, training, validation, and test datasets are mutually exclusive and unknown to each other. The number of epochs is determined using early stopping criteria based on the validation set. We assess the performance of the proposed method using the metrics of True Positive Rate (TPR) and True Negative Rate (TNR). The TPR, also known as sensitivity, quantifies the proportion of actual positive cases (Trojan nodes) that are correctly identified by the model. Conversely, the TNR, or specificity, measures the proportion of actual negative cases (non-Trojan nodes) that are accurately classified. These metrics are calculated using the following equations:

\[
\begin{aligned}
\textit{TPR} &= \frac{\textit{TP}}{\textit{TP} + \textit{FN}} \quad & \textit{TNR} &= \frac{\textit{TN}}{\textit{TN} + \textit{FP}}
\end{aligned}
\]

\subsection{\name~for Different Locality Sizes}
We also experiment the performance of \name~ with different locality sizes ($\mathcal{L}$) for structural features extraction, specifically sizes of $5$, $7$, and $10$. The results, summarized in Table~\ref{tab:localities}, indicate that $\mathcal{L}$ = $7$ yields the highest accuracy in detecting hardware Trojans. Our observation finds that $\mathcal{L}$ = $7$ can be attributed to an optimal balance between context and relevance. A smaller locality size, such as $5$, may not capture sufficient structural information surrounding each node, potentially missing critical patterns associated with Trojan circuits. On the other hand, increasing the locality size to $10$ or more introduces a larger context but also brings in additional irrelevant or noisy information.

\input{fig_latex/explainability_nonTrojans}
\input{Table/Rules}
\subsection{Comparison with the State-of-the-Art}
We further evaluate our proposed framework against several state-of-the-art hardware Trojan detection methods, including VIPR, TrojanSAINT, GNN4Gate, GNN4HT, FAST-GO, R-HTDET, and NHTD. We have also evaluated the performance of several GNN architectures (our own built) on GCN, GCN+JK, GAT, GAT+JK, and XAI-enhanced GAT+JK. As summarized in Table~\ref{tab:sail-ht_main}, XAI-enhanced GAT+JK of \name~outperforms these existing methods by achieving a TPR of $98.47\%$ and a TNR of $98.14\%$. In our approach, we ensure that no benchmarks from the same circuit family are included in either the training or validation datasets. For instance, benchmarks like s38417\_t100, s38417\_t200, and s38417\_t300, which belong to the same circuit family (SCF) of s38417, are excluded from training and validation when evaluating any benchmark from s38417 (shown in Table~\ref{table:data_splitting}). This methodology closely mirrors practical applications where the test data is entirely unknown during the model development phase. Prior works such as GNN4Gate, FAST-GO, R-HTDET and others include benchmarks from the same circuit family in their training sets potentially reduce generalization to new designs. Among them, TrojanSAINT employs a practical training approach, but exhibits a TPR of $82.07\%$ and a TNR of $83.33\%$. As including designs from the same circuit family can potentially enhance model familiarity, NHTD is more effective at correctly identifying non-Trojan nodes, but less capable of detecting actual Trojan instances. 

\subsection{Important features for Trojan nodes}
Figure~\ref{fig:explainability_one} provides a comprehensive illustration of the feature scores conducted using XAI for detecting Trojan nodes. The analysis spans across two distinct benchmarks: s35932\_t200 and s38417\_t200. Each plot represents the top $n=10$ features ranked by their contribution to the model's decision-making process. The aim is to showcase what features are important for the Trojan nodes. $G1$, $G2$ represent gates in the subgraph, with numerical labels denoting their order and values next to them indicate the logic gate type when structural features are gathered using BFS. $Adj-23$ indicates adjacency connections between $G2$ and $G3$. Observation from s38417\_t200: The most influential feature is $G2\_AND$, followed by $G3\_AND$. Connection between $G2$, $G3$ (Adj-23) is significant. 

\subsection{Explainability: Enhanced Findings}

Fig.~\ref{fig:explainability_two} shows the feature distribution analysis which highlights the model's decision-making process for Trojan and non-Trojan nodes (see Fig.~\ref{fig:explainability_two}A) as well as aids in improving its classification accuracy (see Fig.~\ref{fig:explainability_two}B \& Table~\ref{tab:sail-ht_main}). A clear distinction is observed, where Trojan nodes exhibit significantly higher feature values compared to non-Trojan nodes. This sharp difference between the feature scores allows the model to effectively differentiate Trojan nodes from non-Trojan nodes. Fig.~\ref{fig:explainability_two}B delves into finding false negatives within the classification from non-Trojan nodes. In this case, certain non-Trojan nodes show unexpectedly high feature values, almost similar to those of Trojan nodes. Finally, such instances are reclassified as true positives (non-Trojan into Trojan).  
\subsection{Generating Human-Readable Rules}
 The $Captum$-$Explainer$ algorithm helps creating rules from model's decision process. In s35932\_t200 benchmark, the gates $G1$, $G2$, $G3$, $G5$ all are AND gates (Fig.~\ref{fig:explainability_one}) which significantly influences the model's prediction toward identifying a Trojan node. This rule along with other benchmarks' feature scores are translated into human-readable rules in Table~\ref{table:rules}. Additionally, the table highlights two common rules that are applicable across all benchmarks. We deduce that the presence of an AND gate in the local neighborhood plays a pivotal role in steering the prediction toward detecting a Trojan.

\section{Conclusion}
Detecting malicious modifications in digital designs is crucial for ensuring trust and security of modern electronics driving the recent advances in Artificial Intelligence. State-of-the-art hardware Trojans detection solutions suffer from scalability concerns, lack detection precision/stability, and can be a victim of AI hallucinations. To mitigate these concerns, we propose \name~that utilizes Jumping Knowledge mechanism to improve performance and an explainability-guided dynamic post-processing module for minimizing AI hallucinations. \name~outperformed all SOTA Trojan detection frameworks across a large set of benchmarks. Future works will extend \name~for detecting other threats (e.g. side-channel leakage, fault injection vulnerability).        


\bibliographystyle{IEEEtran}
\bibliography{IEEEabrv,sail-ht}


\end{document}

%% file: fig_latex/methodology.tex
\begin{figure*}[]

\includegraphics[width=1\linewidth]{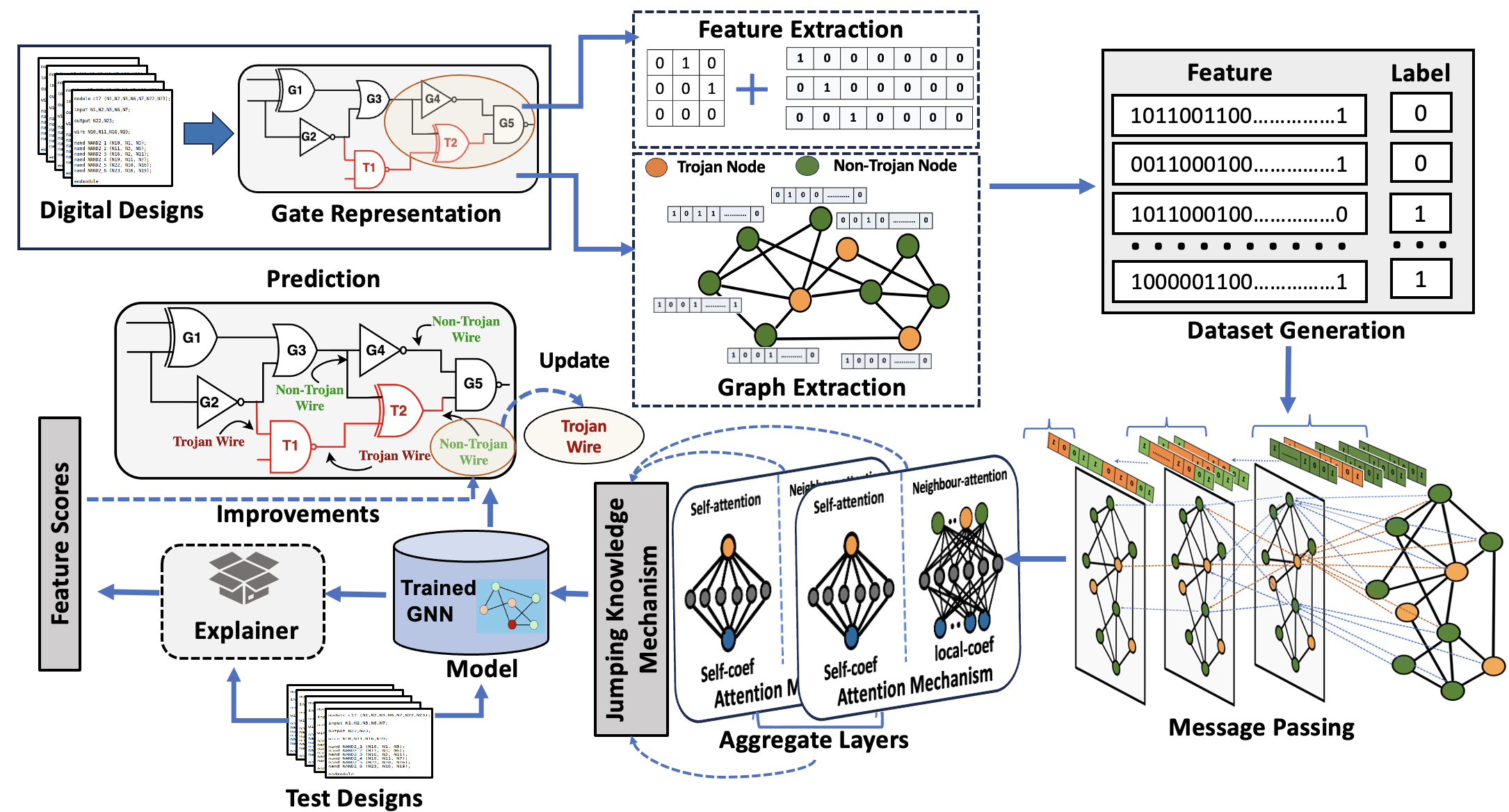}
\caption{The \name~Framework: input/output, feature extraction, graph neural network, and post-processing.   \label{fig:method}}
\vspace{-0.15in}
\end{figure*} 

%% file: Table/Data_Splitting.tex
\begin{table*}[]
\captionsetup{justification=centering}
\caption{Summary of the state-of-the-art Trojan detection frameworks and comparison with \name.}
\label{table:data_splitting}
\renewcommand{\arraystretch}{0.15}
\footnotesize\addtolength{\tabcolsep}{3.3pt}
\begin{threeparttable}
\begin{tabular}{cccccccc}
\toprule
\rowcolor[HTML]{ECF4FF} 
 \begin{tabular}[c]{@{}c@{}}\textbf{Research} \\ \textbf{Study}\end{tabular} & 
 \begin{tabular}[c]{@{}c@{}}\textbf{ML} \\ \textbf{Technique}\end{tabular} & \begin{tabular}[c]{@{}c@{}}\textbf{Data Balancing}\\ \textbf{Technique}\end{tabular} & \begin{tabular}[c]{@{}c@{}}\textbf{Post} \\ \textbf{Processing}\end{tabular} & \begin{tabular}[c]{@{}c@{}}\textbf{Model Training} \\ \textbf{Applicability}\end{tabular} & \begin{tabular}[c]{@{}c@{}}\textbf{No Validation} \\ \textbf{in SCF\tnote{$\dagger$}}\end{tabular} & \begin{tabular}[c]{@{}c@{}}\textbf{No Training} \\ \textbf{in SCF\tnote{$\dagger$}}\end{tabular} & \begin{tabular}[c]{@{}c@{}}\textbf{Model} \\ \textbf{Insight}\end{tabular} \\ 
 
 \midrule
\textbf{VIPR\textsuperscript{*} \cite{gaikwad2023hardware}}                                                                                                               & SVM                                                     & Min Max Scaler                                                     & Yes                                                        & Practical                                                               & N/A                                                             & N/A  & \textcolor[RGB]{190,0,16} {\xmark}                                                             \\ \midrule

\textbf{TrojanSAINT \cite{lashen2023trojansaint}}                                                                                                           & GCN                                                     & Threshold Tuning                                                   & No & Practical                                                               & \textcolor[RGB]{190,0,16} {\xmark}                                                               & \textcolor[RGB]{50,180,50}{\checkmark}       & \textcolor[RGB]{190,0,16} {\xmark}                                                       \\ \midrule
\textbf{XGB \cite{li2020xgboost}}                                                                                                                 & XGB                                                     & Class Weights                                                      & No                                                         & Not Practical                                                           & N/A                                                             & \textcolor[RGB]{190,0,16} {\xmark}    & \textcolor[RGB]{190,0,16} {\xmark}                                                           \\ \midrule
\textbf{GNN4Gate \cite{cheng2022gnn4gate}}                                                                                                             & GCN                                                     & Class Weights                                                      & No                                                          & Not Practical                                                           & \textcolor[RGB]{190,0,16} {\xmark}                                                              & \textcolor[RGB]{190,0,16} {\xmark}   & \textcolor[RGB]{190,0,16} {\xmark}                                                            \\ \midrule
\textbf{GNN4HT \cite{chen2024gnn4ht}}                                                                                                              & GIN                                                     & Data Augmentaion                                                   & No                                                          & Not Practical                                                           &  \textcolor[RGB]{190,0,16} {\xmark}                                                             & \textcolor[RGB]{190,0,16} {\xmark}    & \textcolor[RGB]{190,0,16} {\xmark}                                                        \\ \midrule
\textbf{FAST-GO \cite{imangholi2024fast}}                                                                                                              & GCN                                                     & Class Weights                                                      & No                                                         & Not Practical                                                           & \textcolor[RGB]{190,0,16} {\xmark}                                                              & \textcolor[RGB]{190,0,16} {\xmark}  & \textcolor[RGB]{190,0,16} {\xmark}                                                            \\ \midrule
\textbf{R-HTDET \cite{hasegawa2022r}}                                                   
& GAN                                                     & Oversampling                                                       & No                                                         & Not Practical                                                           & \textcolor[RGB]{190,0,16} {\xmark}                                                              & \textcolor[RGB]{190,0,16} {\xmark}   & \textcolor[RGB]{190,0,16} {\xmark}                                                           \\ \midrule
\textbf{NHTD \cite{hasegawa2023node}}                                                                                                                  & GIN + GAT                                               & Custom Sampling                                                    & No                                                         & Not Practical                                                               & \textcolor[RGB]{190,0,16} {\xmark}                                                              & \textcolor[RGB]{190,0,16} {\xmark}  & \textcolor[RGB]{190,0,16} {\xmark}                                                            \\ \midrule
\textbf{\name}                                                                                                               & GAT + JK                                                & Class Weights                                                      & Yes                                                        & More Practical                                                          & \textcolor[RGB]{50,180,50}{\checkmark}                                                             & \textcolor[RGB]{50,180,50}{\checkmark}     & \textcolor[RGB]{50,180,50}{\checkmark}                                           \\ \midrule             
\end{tabular}
\vspace{0.1pt}
\raggedright \footnotesize{$\dagger$ Refers to Same Circuit Family. For example, rs232\_t1000, rs232\_t1100 belong to same circuit family. VIPR\textsuperscript{*} uses pseudo-self referencing for training.\\}
\end{threeparttable}
\vspace{-0.2in}
\end{table*}

%% file: Algo/datamodel.tex
\setlength{\textfloatsep}{0pt}
\begin{algorithm}[!t]
\DontPrintSemicolon 
\KwIn{$\mathcal{D}, \mathcal{T}_{lb}, \mathcal{L}$}
\KwOut{$\mathbb{M}$}

$G_r \gets$ \textbf{\textit{graphify}}($\mathcal{D}$) \;
$E_d \gets$ \textbf{\textit{edging}}($G_r$) \;
$X\_data, Y\_data \gets \emptyset $; 
$\mathcal{D}_{wires} \gets G_r.wires$ \;
\For{$w$ \textbf{in} $\mathcal{D}_{wires}$}
{
    $S_f \gets$ \textbf{\textit{structural\_features}}($G_r,\mathcal{L},w$)\;
    $X\_data.append(S_f)$ \;
    $Y\_data.append(\mathcal{T}_{lb}[w])$ \;
}
$M_d \gets$ \textbf{\textit{build\_model}}($GAT, JK$) \;
$\mathbb{M} \gets$ \textbf{\textit{training}}($M_d, X\_data$,$Y\_data, E_d$)\;
\Return{$\mathbb{M}$}\;
\caption{Model Generator}
\label{algo:Model Genesis}
\end{algorithm}

%% file: Algo/xai.tex
\setlength{\textfloatsep}{0pt}
\begin{algorithm}[!t]
\DontPrintSemicolon 
\KwIn{$\mathcal{D}, \mathbb{M}, \mathcal{T}_{lb}, \mathcal{L}, n, \mathcal{T}_{h}, \mathcal{N}$}
\KwOut{$\mathbb{TPR}$, $\mathbb{TNR}$}

$G_r \gets$ \textbf{\textit{graphify}}($\mathcal{D}$) \;
$E_d \gets$ \textbf{\textit{edging}}($G_r$) \;
$S_f \gets$ \textbf{\textit{structural\_features}}($G_r,\mathcal{L},G_r.wires$)\;
$P_0, P_1 \gets$ \textbf{\textit{prediction}}($\mathbb{M}, S_f, E_d, G_r.wires$)\;
$x, y \gets$ \textbf{\textit{len}}($P_0$),  \textbf{\textit{len}}($P_1$)\;
$F_0[x][n], F_1[y][n] \gets$ \textbf{\textit{explainer}}($\mathbb{M}, P_0, P_1, n$)\;
\For{$j$ \textbf{in} $range(n)$}
{
    $\mathcal{V}[0][j] \gets$ \textbf{\textit{sum\_avg}}($F_0[i][j]:i \gets range(len(F_0))$)\;
    $\mathcal{K}[0][j] \gets$ \textbf{\textit{sum\_avg}}($F_1[i][j]:i \gets range(len(F_1))$)\;
}
\For{$i$ \textbf{in} $range(F_0)$}
{
    $\mathcal{C}_{1} \gets sum(F_0[i][n]) \ge (sum(\mathcal{V}[0][n]) * \mathcal{N})$ \;
    $\mathcal{C}_{2} \gets |sum(F_0[i][n]) - sum(\mathcal{K}[0][n])| \ge \mathcal{T}_{h}$ \;
    \If{$\mathcal{C}_1$ $or$ $\mathcal{C}_2$} 
    {
        \textbf{\textit{switch\_pred}}($P_0, P_1$)\;
    }
}
$\mathbb{TPR}$, $\mathbb{TNR} \gets$ \textbf{\textit{compute\_metrics}}($P_0, P_1, \mathcal{T}_{lb}$) \;
\Return{$\mathbb{TPR}$, $\mathbb{TNR}$}\;
\caption{XAI Development}
\label{algo:XAI Development}
\end{algorithm}

%% file: Table/SAIL-HT_Locality.tex
\begin{table}[]
\centering
\captionsetup{justification=centering}
\caption{\name~for different localities.}
\label{tab:localities}
\renewcommand{\arraystretch}{1}
\footnotesize\addtolength{\tabcolsep}{1pt}
\begin{tabular}{|l|cc|cc|cc|}
\hline
\rowcolor[HTML]{ECF4FF} 
\multicolumn{1}{|c|}{\cellcolor[HTML]{ECF4FF}} &
  \multicolumn{2}{c|}{\cellcolor[HTML]{ECF4FF}\textbf{Locality = 5}} &
  \multicolumn{2}{c|}{\cellcolor[HTML]{ECF4FF}\textbf{Locality = 7}} &
  \multicolumn{2}{c|}{\cellcolor[HTML]{ECF4FF}\textbf{Locality = 10}} \\ \cline{2-7} 
\rowcolor[HTML]{EFEFEF} 
\multicolumn{1}{|c|}{\multirow{-2}{*}{\cellcolor[HTML]{ECF4FF}\textbf{Benchmarks}}} &
  \multicolumn{1}{c|}{\cellcolor[HTML]{EFEFEF}\textbf{TPR}} &
  \textbf{TNR} &
  \multicolumn{1}{c|}{\cellcolor[HTML]{EFEFEF}\textbf{TPR}} &
  \textbf{TNR} &
  \multicolumn{1}{c|}{\cellcolor[HTML]{EFEFEF}\textbf{TPR}} &
  \textbf{TNR} \\ \hline
\textbf{rs232\_t1000} & \multicolumn{1}{c|}{40.00}  & 100 & \multicolumn{1}{c|}{100} & 96.63  & \multicolumn{1}{c|}{100} & 96.03  \\
\rowcolor[HTML]{FFFFFF} 
\textbf{rs232\_t1100} &
  \multicolumn{1}{c|}{\cellcolor[HTML]{FFFFFF}36.36} &
  79 &
  \multicolumn{1}{c|}{\cellcolor[HTML]{FFFFFF}100} &
  94.88 &
  \multicolumn{1}{c|}{\cellcolor[HTML]{FFFFFF}100} &
  93.67 \\
\textbf{rs232\_t1200} & \multicolumn{1}{c|}{100} & 72.04  & \multicolumn{1}{c|}{100} & 96.31  & \multicolumn{1}{c|}{100} & 94.41  \\
\textbf{rs232\_t1300} & \multicolumn{1}{c|}{57.14}  & 92.13  & \multicolumn{1}{c|}{100} & 96.32  & \multicolumn{1}{c|}{100} & 98.36  \\
\textbf{rs232\_t1400} & \multicolumn{1}{c|}{100} & 72.70  & \multicolumn{1}{c|}{100} & 95.93  & \multicolumn{1}{c|}{100} & 92.46  \\
\textbf{rs232\_t1500} & \multicolumn{1}{c|}{63.64}  & 78.74  & \multicolumn{1}{c|}{100} & 94.61  & \multicolumn{1}{c|}{100} & 92.69  \\
\textbf{rs232\_t1600} & \multicolumn{1}{c|}{42.86}  & 93.07  & \multicolumn{1}{c|}{100} & 97.66  & \multicolumn{1}{c|}{100} & 99.67  \\
\textbf{rs232\_t1700} & \multicolumn{1}{c|}{70}     & 92.74  & \multicolumn{1}{c|}{100} & 98.33  & \multicolumn{1}{c|}{100} & 99.67  \\ \hline
\textbf{s15850\_t100} & \multicolumn{1}{c|}{100}    & 98.76  & \multicolumn{1}{c|}{100} & 99.70  & \multicolumn{1}{c|}{100} & 99.43  \\ \hline
\textbf{s38417\_t100}  & \multicolumn{1}{c|}{91.67}  & 99.78  & \multicolumn{1}{c|}{100} & 100 & \multicolumn{1}{c|}{91.67}  & 100 \\
\textbf{s38417\_t200}  & \multicolumn{1}{c|}{66.67}  & 99.93  & \multicolumn{1}{c|}{93.33}  & 93.33  & \multicolumn{1}{c|}{73.33}  & 100\\
\textbf{s38417\_t300}  & \multicolumn{1}{c|}{100} & 99.93  & \multicolumn{1}{c|}{100} & 100 & \multicolumn{1}{c|}{94.12}  & 100 \\ \hline
\textbf{s38584\_t100} &
  \multicolumn{1}{c|}{38.89} &
  \multicolumn{1}{c|}{99.89} &
  \multicolumn{1}{c|}{88.89} &
  \multicolumn{1}{c|}{99.98} &
  \multicolumn{1}{c|}{55.56} &
  \multicolumn{1}{c|}{100} \\ \hline
\textbf{s35932 t100}  & \multicolumn{1}{c|}{86.67}  & 98.68  & \multicolumn{1}{c|}{93.33}  & 100 & \multicolumn{1}{c|}{86.67}  & 100 \\
\textbf{s35932 t200}  & \multicolumn{1}{c|}{66.67}  & 98.71  & \multicolumn{1}{c|}{100} & 99.91  & \multicolumn{1}{c|}{75.00}  & 100 \\
\textbf{s35932 t300}  & \multicolumn{1}{c|}{66.67}  & 98.86  & \multicolumn{1}{c|}{100} & 99.96  & \multicolumn{1}{c|}{75}  & 99.97  \\ \hline
\rowcolor[HTML]{F2FAED} 
\multicolumn{1}{|c|}{\cellcolor[HTML]{F2FAED}\textbf{Average}} &
  \multicolumn{1}{c|}{\cellcolor[HTML]{F2FAED}\textbf{70.45}} &
  \textbf{92.18} &
  \multicolumn{1}{c|}{\cellcolor[HTML]{F2FAED}\textbf{98.47}} &
  \textbf{98.14} &
  \multicolumn{1}{c|}{\cellcolor[HTML]{F2FAED}\textbf{90.71}} &
  \textbf{97.88} \\ \hline
\end{tabular}
\end{table}

%% file: Table/SAIL-HT_main.tex
\begin{table*}[!htbp]
\centering
\captionsetup{justification=centering}
\caption{Quantitative comparison between \name~ and  other state-of-the-art frameworks. \name~outperforms all.}

\label{tab:sail-ht_main}
\renewcommand{\arraystretch}{1.1}
\footnotesize\addtolength{\tabcolsep}{-2.5pt}
\begin{tabular}{|c|cc|cc|cc|cc|cc|cc|cc|cccc|}
\hline
\rowcolor[HTML]{ECF4FF} 
\cellcolor[HTML]{ECF4FF} &
  \multicolumn{2}{c|}{\cellcolor[HTML]{ECF4FF}} &
  \multicolumn{2}{c|}{\cellcolor[HTML]{ECF4FF}} &
  \multicolumn{2}{c|}{\cellcolor[HTML]{ECF4FF}} &
  \multicolumn{2}{c|}{\cellcolor[HTML]{ECF4FF}} &
  \multicolumn{2}{c|}{\cellcolor[HTML]{ECF4FF}} &
  \multicolumn{2}{c|}{\cellcolor[HTML]{ECF4FF}} &
  \multicolumn{2}{c|}{\cellcolor[HTML]{ECF4FF}} &
  \multicolumn{4}{c|}{\cellcolor[HTML]{ECF4FF}\textbf{\name}} \\ \cline{16-19} 
\rowcolor[HTML]{ECF4FF} 
\cellcolor[HTML]{ECF4FF} &
  \multicolumn{2}{c|}{\multirow{-2}{*}{\cellcolor[HTML]{ECF4FF}\textbf{\begin{tabular}[c]{@{}c@{}}VIPR\\ {\cite{gaikwad2023hardware}}\end{tabular}}}} &
  \multicolumn{2}{c|}{\multirow{-2}{*}{\cellcolor[HTML]{ECF4FF}\textbf{\begin{tabular}[c]{@{}c@{}}TrojanSAINT\\ {\cite{lashen2023trojansaint}}\end{tabular}}}} &
  \multicolumn{2}{c|}{\multirow{-2}{*}{\cellcolor[HTML]{ECF4FF}\textbf{\begin{tabular}[c]{@{}c@{}}GNN4Gate\\ {\cite{cheng2022gnn4gate}}\end{tabular}}}} &
  \multicolumn{2}{c|}{\multirow{-2}{*}{\cellcolor[HTML]{ECF4FF}\textbf{\begin{tabular}[c]{@{}c@{}}GNN4HT\\ {\cite{chen2024gnn4ht}}\end{tabular}}}} &
  \multicolumn{2}{c|}{\multirow{-2}{*}{\cellcolor[HTML]{ECF4FF}\textbf{\begin{tabular}[c]{@{}c@{}}FAST-GO\\ {\cite{imangholi2024fast}}\end{tabular}}}} &
  \multicolumn{2}{c|}{\multirow{-2}{*}{\cellcolor[HTML]{ECF4FF}\textbf{\begin{tabular}[c]{@{}c@{}}R-HTDET\\ {\cite{hasegawa2022r}}\end{tabular}}}} &
  \multicolumn{2}{c|}{\multirow{-2}{*}{\cellcolor[HTML]{ECF4FF}\textbf{\begin{tabular}[c]{@{}c@{}}NHTD\\ {\cite{hasegawa2023node}}\end{tabular}}}} &
  \multicolumn{2}{c|}{\cellcolor[HTML]{ECF4FF}\textbf{GAT+JK}} &
  \multicolumn{2}{c|}{\cellcolor[HTML]{ECF4FF}\textbf{XAI-GAT+JK}} \\ \cline{2-19} 
\rowcolor[HTML]{EFEFEF} 
\multirow{-3}{*}{\cellcolor[HTML]{ECF4FF}\textbf{Benchmarks}} &
  \textbf{TPR} &
  \textbf{TNR} &
  \textbf{TPR} &
  \textbf{TNR} &
  \textbf{TPR} &
  \textbf{TNR} &
  \textbf{TPR} &
  \textbf{TNR} &
  \textbf{TPR} &
  \textbf{TNR} &
  \textbf{TPR} &
  \textbf{TNR} &
  \textbf{TPR} &
  \textbf{TNR} &
  \textbf{TPR} &
  \multicolumn{1}{c|}{\cellcolor[HTML]{EFEFEF}\textbf{TNR}} &
  \textbf{TPR} &
  \textbf{TNR} \\ \hline
\rowcolor[HTML]{FFFFFF} 
\textbf{rs232\_t1000} &
  100 &
  98.68 &
  100 &
  60 &
  100 &
  100 &
  100 &
  99 &
  100 &
  84.36 &
  100 &
  94.30 &
  100 &
  100 &
  100 &
  \multicolumn{1}{c|}{\cellcolor[HTML]{FFFFFF}96.69} &
  \cellcolor[HTML]{FFFFFF}100 &
  96.63 \\
\rowcolor[HTML]{FFFFFF} 
\textbf{rs232\_t1100} &
  50 &
  98.71 &
  92 &
  68 &
  100 &
  100 &
  100 &
  99 &
  100 &
  84.30 &
  100 &
  93.30 &
  100 &
  99.66 &
  100 &
  \multicolumn{1}{c|}{\cellcolor[HTML]{FFFFFF}95} &
  \cellcolor[HTML]{FFFFFF}100 &
  94.88 \\
\rowcolor[HTML]{FFFFFF} 
\textbf{rs232\_t1200} &
  76.47 &
  99.67 &
  41 &
  80 &
  100 &
  100 &
  100 &
  99 &
  100 &
  84.77 &
  93.20 &
  96.50 &
  100 &
  99.66 &
  \cellcolor[HTML]{FFFFFF}100 &
  \multicolumn{1}{c|}{\cellcolor[HTML]{FFFFFF}96.38} &
  \cellcolor[HTML]{FFFFFF}100 &
  96.31 \\
\rowcolor[HTML]{FFFFFF} 
\textbf{rs232\_t1300} &
  100 &
  99.67 &
  100 &
  74 &
  100 &
  100 &
  100 &
  99 &
  100 &
  83.13 &
  100 &
  94.40 &
  100 &
  100 &
  \cellcolor[HTML]{FFFFFF}100 &
  \multicolumn{1}{c|}{\cellcolor[HTML]{FFFFFF}96.39} &
  \cellcolor[HTML]{FFFFFF}100 &
  96.32 \\
\rowcolor[HTML]{FFFFFF} 
\textbf{rs232\_t1400} &
  92.31 &
  98 &
  92 &
  50 &
  100 &
  100 &
  100 &
  99.50 &
  100 &
  84.36 &
  100 &
  97.80 &
  100 &
  99.66 &
  \cellcolor[HTML]{FFFFFF}100 &
  \multicolumn{1}{c|}{\cellcolor[HTML]{FFFFFF}96} &
  \cellcolor[HTML]{FFFFFF}100 &
  95.93 \\
\rowcolor[HTML]{FFFFFF} 
\textbf{rs232\_t1500} &
  91.67 &
  99.67 &
  71 &
  82 &
  100 &
  100 &
  100 &
  99 &
  100 &
  84.77 &
  98 &
  94.30 &
  100 &
  100 &
  \cellcolor[HTML]{FFFFFF}100 &
  \multicolumn{1}{c|}{\cellcolor[HTML]{FFFFFF}94.68} &
  \cellcolor[HTML]{FFFFFF}100 &
  94.61 \\
\rowcolor[HTML]{FFFFFF} 
\textbf{rs232\_t1600} &
  100 &
  99.67 &
  73 &
  57 &
  100 &
  100 &
  100 &
  100 &
  100 &
  84.02 &
  97.40 &
  92.50 &
  69.23 &
  100 &
  \cellcolor[HTML]{FFFFFF}100 &
  \multicolumn{1}{c|}{\cellcolor[HTML]{FFFFFF}97.69} &
  \cellcolor[HTML]{FFFFFF}100 &
  97.66 \\
\rowcolor[HTML]{FFFFFF} 
\textbf{rs232\_t1700} &
  100 &
  100 &
  - &
  - &
  - &
  - &
  - &
  - &
  - &
  - &
  - &
  - &
  - &
  - &
  \cellcolor[HTML]{FFFFFF}100 &
  \multicolumn{1}{c|}{\cellcolor[HTML]{FFFFFF}98.35} &
  \cellcolor[HTML]{FFFFFF}100 &
  98.33 \\ \hline
\rowcolor[HTML]{FFFFFF} 
\textbf{s15850\_t100} &
  - &
  - &
  35 &
  97 &
  - &
  - &
  59.26 &
  90.35 &
  100 &
  80.93 &
  60.50 &
  90.10 &
  100 &
  100 &
  100 &
  \multicolumn{1}{c|}{\cellcolor[HTML]{FFFFFF}99.70} &
  \cellcolor[HTML]{FFFFFF}100 &
  99.70 \\ \hline
\rowcolor[HTML]{FFFFFF} 
\textbf{s38417\_t100} &
  100 &
  99.91 &
  92 &
  92 &
  100 &
  99.92 &
  100 &
  95.62 &
  91.67 &
  80.15 &
  77.30 &
  99.90 &
  66.67 &
  100 &
  100 &
  \multicolumn{1}{c|}{\cellcolor[HTML]{FFFFFF}99.99} &
  \cellcolor[HTML]{FFFFFF}100 &
  99.99 \\
\rowcolor[HTML]{FFFFFF} 
\textbf{s38417\_t200} &
  100 &
  99.98 &
  40 &
  99 &
  80 &
  99.85 &
  100 &
  97.24 &
  100 &
  80.20 &
  84 &
  98.70 &
  100 &
  100 &
  80 &
  \multicolumn{1}{c|}{\cellcolor[HTML]{FFFFFF}99.90} &
  93.33 &
  99.98 \\
\rowcolor[HTML]{FFFFFF} 
\textbf{s38417\_t300} &
  - &
  - &
  98 &
  96 &
  100 &
  99.46 &
  53.33 &
  95.13 &
  100 &
  80.21 &
  83 &
  99.90 &
  93.33 &
  99.98 &
  94.12 &
  \multicolumn{1}{c|}{\cellcolor[HTML]{FFFFFF}100} &
  100 &
  100.00 \\ \hline
\rowcolor[HTML]{FFFFFF} 
\textbf{s38584\_t100} &
  - &
  - &
  100 &
  95 &
  - &
  - &
  88.89 &
  87.72 &
  68.42 &
  80.13 &
  - &
  - &
  33.33 &
  100 &
  72.22 &
  \multicolumn{1}{c|}{\cellcolor[HTML]{FFFFFF}99.98} &
  88.89 &
  99.98 \\ \hline
\rowcolor[HTML]{FFFFFF} 
\textbf{s35932\_t100} &
  - &
  - &
  100 &
  100 &
  93.33 &
  99.98 &
  93.33 &
  96.19 &
  100 &
  80.19 &
  80 &
  75.20 &
  80 &
  100 &
  80 &
  \multicolumn{1}{c|}{\cellcolor[HTML]{FFFFFF}100} &
  93.33 &
  100 \\
\rowcolor[HTML]{FFFFFF} 
\textbf{s35932\_t200} &
  - &
  - &
  100 &
  100 &
  68.75 &
  100 &
  93.75 &
  91.83 &
  70.59 &
  80.13 &
  36.40 &
  99.90 &
  73.33 &
  100 &
  83.33 &
  \multicolumn{1}{c|}{\cellcolor[HTML]{FFFFFF}100} &
  100 &
  99.91 \\
\rowcolor[HTML]{FFFFFF} 
\textbf{s35932\_t300} &
  - &
  - &
  97 &
  100 &
  33.33 &
  100 &
  100 &
  97.97 &
  100 &
  80.45 &
  88 &
  100 &
  80.77 &
  99.95 &
  83.33 &
  \multicolumn{1}{c|}{\cellcolor[HTML]{FFFFFF}100} &
  100 &
  99.96 \\ \hline
\rowcolor[HTML]{F2FAED} 
\textbf{Average} &
  \textbf{91.04} &
  \textbf{99.40} &
  \textbf{82.07} &
  \textbf{83.33} &
  \textbf{90.42} &
  \textbf{99.94} &
  \textbf{92.57} &
  \textbf{96.44} &
  \textbf{95.38} &
  \textbf{82.14} &
  \textbf{85.56} &
  \textbf{94.77} &
  \textbf{86.44} &
  \textbf{99.93} &
  \multicolumn{1}{l}{\cellcolor[HTML]{F2FAED}\textbf{93.31}} &
  \multicolumn{1}{c|}{\cellcolor[HTML]{F2FAED}\textbf{98.18}} &
  \textbf{98.47} &
  \textbf{98.14}\\ \hline
\end{tabular}
\vspace{-0.1in}
\end{table*}

%% file: fig_latex/explainability_Trojans.tex
\begin{figure*}[]
\centering

\includegraphics[width=0.85\linewidth]{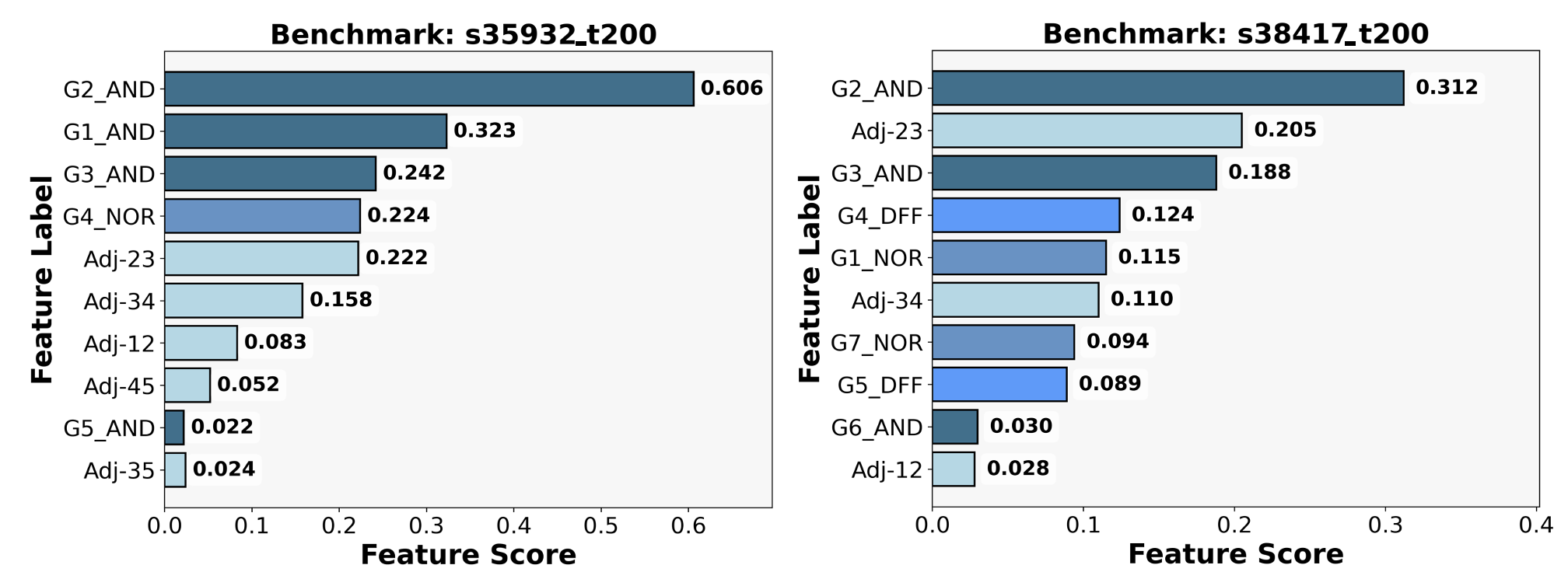}
\caption{Illustration of \name~model's decision-making process for Trojan nodes. Two distinct benchmarks are shown here.  \label{fig:explainability_one}}

\vspace{-0.2in}
\end{figure*}

%% file: fig_latex/explainability_nonTrojans.tex



\begin{figure*}[]
\centering

\includegraphics[width=\linewidth]{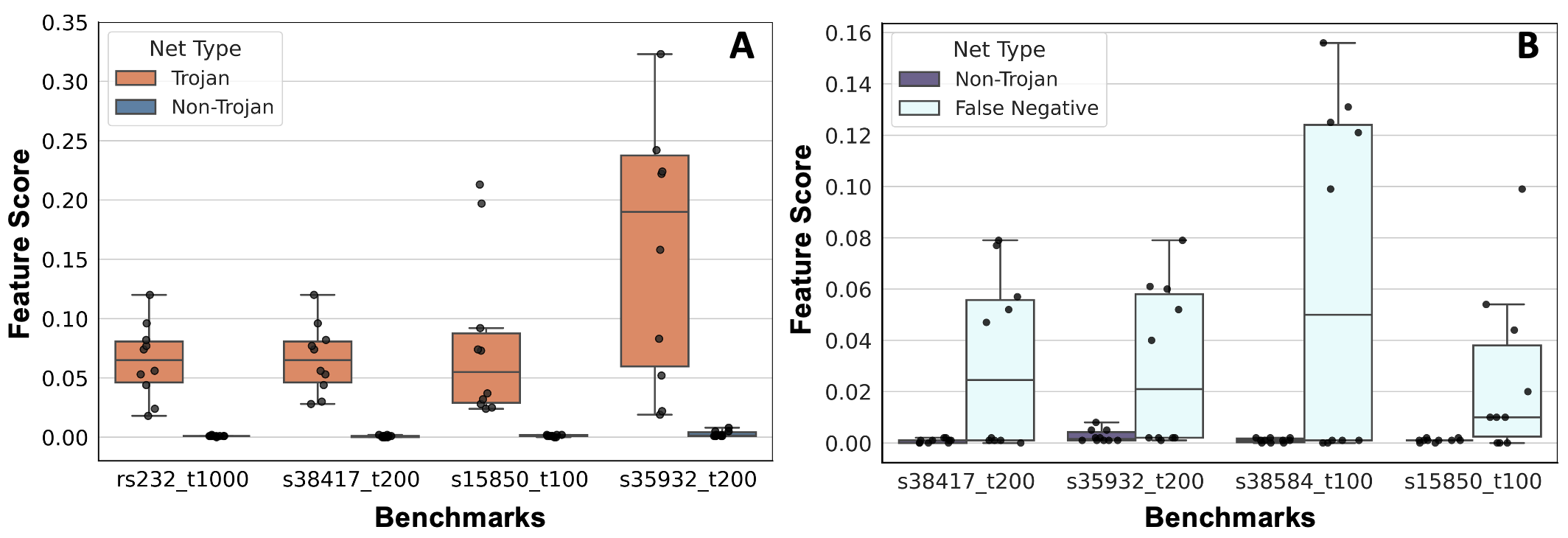}
\caption{\textbf{A}: Comparison of feature scores between Trojan and non-Trojan nodes - Trojan nodes exhibit significantly higher feature scores compared to non-Trojan nodes; \textbf{B}: Analysis of feature scores for non-Trojan nodes - Certain non-Trojan nodes generate high feature values, indicating false negatives in the classification.\label{fig:explainability_two}}

\vspace{-0.1in}
\end{figure*}

%% file: Table/Rules.tex
\begin{table*}[]
\centering
\captionsetup{justification=centering}
\caption{XAI generated human-interpretable rules for Trojan detection.} 
\label{table:rules}
\renewcommand{\arraystretch}{1.2}
\footnotesize\addtolength{\tabcolsep}{5pt}
\begin{tabular}{|
>{\columncolor[HTML]{EFEFEF}}c |c|}
\hline
\cellcolor[HTML]{ECF4FF}\textbf{Benchmark} & \cellcolor[HTML]{ECF4FF}\textbf{What the AI Model learned to classify Trojan}                                                                                                                                                                     \\ \hline
\textbf{s38417\_t200}                      & \begin{tabular}[c]{@{}c@{}} If {[}G2 = AND, G3 = AND \&\& G4, G5 = D Flip Flop, G1 = NOR \&\& G2, G3 are connected \\ \&\& G3, G4 are connected{]} is the locality of a net then there is a higher possibility that it can be a Trojan.\end{tabular}                                    \\ \hline
\textbf{rs232-t1000}                       & \begin{tabular}[c]{@{}c@{}} If {[}G2 = OR, G3 = OR \&\& G2,G3 connected to each other  \&\& G4 = AND, G7 is NONE gate \&\& \\ G3,G4 are also connected {]}  is the locality of a net then there is a higher possibility that it can be a Trojan. \end{tabular}                              \\ \hline
\textbf{s15850-t100}                       & \begin{tabular}[c]{@{}c@{}} If {[}G1, G2, G3, G4 all are AND gate \&\& G2 is connected with G3 \&\& G1 is connected to G2  \&\& G5 = D Flip Flop \\ \&\& G4, G5 are connected{]} is the locality of a net then there is a higher possibility that it can be a Trojan.\end{tabular}       \\ \hline
\textbf{s38584\_t100}                      & \begin{tabular}[c]{@{}c@{}} If {[}G6, G7 are NONE gate \&\& G1, G2, G3  all are AND gate \&\&  G1,G2, \\G3  are connected among it-selves{]} is the locality of a net then there is a higher possibility that it can be a Trojan.\end{tabular}                                              \\ \hline
\textbf{s35932-t200}                       & \begin{tabular}[c]{@{}c@{}} If {[}G1,G2,G3,G5 all are AND gate \&\& G4 = NOR \&\& G1,G2 are connected \&\& G2,G3 are connected  \&\& G3,G4 \\are connected  \&\& G4,G5 are connected{]} is the locality of a net then there is a higher possibility that it can be a Trojan.\end{tabular} \\ \hline
\textbf{Any}                               & Trojan node's feature importance score is significantly higher than that of any other nodes.                                                                                                                                                                                                                             \\ \hline
\textbf{Any}                               & Trojan node mostly exists on a locale where it is surrounded by adjacent and connected AND gates.                                                                                                                                                                                                                               \\ \hline
\end{tabular}
\vspace{-0.2in}
\end{table*}